\documentclass[reprint, aps,amssymb,amsmath,showpacs,superscriptaddress] {revtex4-2}%
\usepackage{graphicx}% Include figure files
\usepackage{dcolumn}% Align table columns on decimal point
\usepackage{bm}% bold math
\usepackage{siunitx}%
\usepackage[version=4]{mhchem}
\usepackage{xspace}
\usepackage{placeins}
\usepackage{lineno}
\usepackage{booktabs}
\usepackage{float}
\newcommand*{\pTlead}{\ensuremath{p_\mathrm{T,jet}^\mathrm{lead}}\xspace}

\newcommand*{\pTrawlead}{\ensuremath{p_\mathrm{T,jet}^\mathrm{raw,lead}}\xspace}
\newcommand*{\pTrawsub}{\ensuremath{p_\mathrm{T,jet}^\mathrm{raw,sub}}\xspace}
\newcommand*{\pTrawrecoil}{\ensuremath{p_\mathrm{T,jet}^\mathrm{raw,recoil}}\xspace}

\newcommand*{\ms}[1]{\ensuremath{#1}\xspace}

\newcommand*{\RpdA}{\ms{R_{p/d+\mathrm{A}}^\mathrm{jet}}}
 
\newcommand*{\RpPb}{\ms{R_{p+\mathrm{Pb}}^\mathrm{jet}}} 
\newcommand*{\RpPbjet}{\ms{R_{p+\mathrm{Pb}}^\mathrm{jet}}}
\newcommand*{\AJ}{\ms{A_\mathrm{J}}}

\newcommand*{\Dphi}{\ensuremath{\Delta\phi}\xspace}
\newcommand*{\EAbbc}{\ensuremath{\mathrm{EA}_\mathrm{BBC}}\xspace}
\newcommand*{\ETtrig}{\ms{E_\mathrm{T}^\mathrm{trig}}}
\newcommand*{\ET}{\ms{E_\mathrm{T}}}

\newcommand*{\Rjet} {\ms{R_\mathrm{jet}}}
\newcommand*{\absDphi}{\ensuremath{|\Delta\phi|}\xspace}
\newcommand*{\antikT}{anti-\ensuremath{k_\mathrm{T}}\xspace}

\newcommand*{\pp}{\ms{p\!+\!p\!\!}\xspace}
\newcommand*{\pT}{\ms{p_\mathrm{T}}}
\newcommand*{\sNNcc}{\ensuremath{\sqrt{s_\mathrm{NN}}=\SI{200}{GeV}}\xspace}
\newcommand*{\xp}{\ms{x_\mathrm{p}}}
\newcommand*{\GeantThree}{\texttt{GEANT3}\xspace}
\newcommand*{\fig}[1]{Fig.~\ref{fig:#1}\xspace}
\newcommand*{\PythiaSix}{\texttt{PYTHIA6}\xspace}
\newcommand*{\PythiaEight}{\texttt{PYTHIA8}\xspace}
\newcommand*{\pAu}{$p$+Au\xspace}

\begin{document}
\preprint{APS/123-QED}
\title{
Correlations of event activity with hard and soft processes in $p$ + Au collisions at \sNNcc at STAR
}

\affiliation{Abilene Christian University, Abilene, Texas   79699}
\affiliation{AGH University of Krakow, FPACS, Cracow 30-059, Poland}
\affiliation{Argonne National Laboratory, Argonne, Illinois 60439}
\affiliation{American University in Cairo, New Cairo 11835, Egypt}
\affiliation{Ball State University, Muncie, Indiana, 47306}
\affiliation{Brookhaven National Laboratory, Upton, New York 11973}
\affiliation{University of Calabria \& INFN-Cosenza, Rende 87036, Italy}
\affiliation{University of California, Berkeley, California 94720}
\affiliation{University of California, Davis, California 95616}
\affiliation{University of California, Los Angeles, California 90095}
\affiliation{University of California, Riverside, California 92521}
\affiliation{Central China Normal University, Wuhan, Hubei 430079 }
\affiliation{University of Illinois at Chicago, Chicago, Illinois 60607}
\affiliation{Chongqing University, Chongqing, 401331}
\affiliation{Creighton University, Omaha, Nebraska 68178}
\affiliation{Czech Technical University in Prague, FNSPE, Prague 115 19, Czech Republic}
\affiliation{Technische Universit\"at Darmstadt, Darmstadt 64289, Germany}
\affiliation{National Institute of Technology Durgapur, Durgapur - 713209, India}
\affiliation{ELTE E\"otv\"os Lor\'and University, Budapest, Hungary H-1117}
\affiliation{Frankfurt Institute for Advanced Studies FIAS, Frankfurt 60438, Germany}
\affiliation{Fudan University, Shanghai, 200433 }
\affiliation{Guangxi Normal University, Guilin, 541004}
\affiliation{University of Heidelberg, Heidelberg 69120, Germany }
\affiliation{University of Houston, Houston, Texas 77204}
\affiliation{Huzhou University, Huzhou, Zhejiang  313000}
\affiliation{Indian Institute of Science Education and Research (IISER), Berhampur 760010 , India}
\affiliation{Indian Institute of Science Education and Research (IISER) Tirupati, Tirupati 517507, India}
\affiliation{Indian Institute Technology, Patna, Bihar 801106, India}
\affiliation{Indiana University, Bloomington, Indiana 47408}
\affiliation{Institute of Modern Physics, Chinese Academy of Sciences, Lanzhou, Gansu 730000 }
\affiliation{University of Jammu, Jammu 180001, India}
\affiliation{Kent State University, Kent, Ohio 44242}
\affiliation{University of Kentucky, Lexington, Kentucky 40506-0055}
\affiliation{Lawrence Berkeley National Laboratory, Berkeley, California 94720}
\affiliation{Lehigh University, Bethlehem, Pennsylvania 18015}
\affiliation{Max-Planck-Institut f\"ur Physik, Munich 80805, Germany}
\affiliation{Michigan State University, East Lansing, Michigan 48824}
\affiliation{National Institute of Science Education and Research, HBNI, Jatni 752050, India}
\affiliation{National Cheng Kung University, Tainan 70101 }
\affiliation{Nuclear Physics Institute of the CAS, Rez 250 68, Czech Republic}
\affiliation{The Ohio State University, Columbus, Ohio 43210}
\affiliation{Institute of Nuclear Physics PAN, Cracow 31-342, Poland}
\affiliation{Panjab University, Chandigarh 160014, India}
\affiliation{Purdue University, West Lafayette, Indiana 47907}
\affiliation{Rice University, Houston, Texas 77251}
\affiliation{Rutgers University, Piscataway, New Jersey 08854}
\affiliation{University of Science and Technology of China, Hefei, Anhui 230026}
\affiliation{South China Normal University, Guangzhou, Guangdong 510631}
\affiliation{Sejong University, Seoul, 05006, South Korea}
\affiliation{Shandong University, Qingdao, Shandong 266237}
\affiliation{Shanghai Institute of Applied Physics, Chinese Academy of Sciences, Shanghai 201800}
\affiliation{Southern Connecticut State University, New Haven, Connecticut 06515}
\affiliation{State University of New York, Stony Brook, New York 11794}
\affiliation{Instituto de Alta Investigaci\'on, Universidad de Tarapac\'a, Arica 1000000, Chile}
\affiliation{Temple University, Philadelphia, Pennsylvania 19122}
\affiliation{Texas A\&M University, College Station, Texas 77843}
\affiliation{University of Texas, Austin, Texas 78712}
\affiliation{Tsinghua University, Beijing 100084}
\affiliation{University of Tsukuba, Tsukuba, Ibaraki 305-8571, Japan}
\affiliation{University of Chinese Academy of Sciences, Beijing, 101408}
\affiliation{United States Naval Academy, Annapolis, Maryland 21402}
\affiliation{Valparaiso University, Valparaiso, Indiana 46383}
\affiliation{Variable Energy Cyclotron Centre, Kolkata 700064, India}
\affiliation{Warsaw University of Technology, Warsaw 00-661, Poland}
\affiliation{Wayne State University, Detroit, Michigan 48201}
\affiliation{Wuhan University of Science and Technology, Wuhan, Hubei 430065}
\affiliation{Yale University, New Haven, Connecticut 06520}

\author{M.~I.~Abdulhamid}\affiliation{American University in Cairo, New Cairo 11835, Egypt}
\author{B.~E.~Aboona}\affiliation{Texas A\&M University, College Station, Texas 77843}
\author{J.~Adam}\affiliation{Czech Technical University in Prague, FNSPE, Prague 115 19, Czech Republic}
\author{L.~Adamczyk}\affiliation{AGH University of Krakow, FPACS, Cracow 30-059, Poland}
\author{J.~R.~Adams}\affiliation{The Ohio State University, Columbus, Ohio 43210}
\author{I.~Aggarwal}\affiliation{Panjab University, Chandigarh 160014, India}
\author{M.~M.~Aggarwal}\affiliation{Panjab University, Chandigarh 160014, India}
\author{Z.~Ahammed}\affiliation{Variable Energy Cyclotron Centre, Kolkata 700064, India}
\author{E.~C.~Aschenauer}\affiliation{Brookhaven National Laboratory, Upton, New York 11973}
\author{S.~Aslam}\affiliation{Indian Institute Technology, Patna, Bihar 801106, India}
% ...

\collaboration{STAR Collaboration}\noaffiliation

%\affiliation{%
 %STAR Authors' affiliations }
 
%\collaboration{STAR Collaboration}%\noaffiliation

\date{\today}% It is always \today, today, but any date may be explicitly specified

\begin{abstract} 

\noindent With the STAR experiment at the BNL Relativisitc Heavy Ion Collider, we characterize \sNNcc \pAu collisions by event activity (EA) measured within the pseudorapidity range $\eta\in[-5,-3.4]$ in the Au-going direction and report correlations between this EA and hard- and soft-scale particle production at midrapidity ($\eta\in[-1,1]$). At the soft scale, charged particle production in low-EA $p$+Au collisions is comparable to that in \pp collisions and increases monotonically with increasing EA. At the hard scale, we report measurements of high transverse momentum (\pT) jets in events of different EAs. In contrast with the soft particle production, high-\pT particle production and EA are found to be inversely related. To investigate whether this is a signal of jet quenching in high-EA events, we also report ratios of \pT imbalance and azimuthal separation of dijets in high- and low-EA events. Within our measurement precision, no significant differences are observed, disfavoring the presence of jet quenching in the highest 30\% EA $p$+Au collisions at \sNNcc.
\end{abstract}
\maketitle

%%%%%%%%%%%%%%%%%%%%%%%%%%%%%%%%%%%%%%%
% Section: Introduction               %
%%%%%%%%%%%%%%%%%%%%%%%%%%%%%%%%%%%%%%%
\section{Introduction}\label{sec:Introduction}

Jets are algorithmically clustered groups of nearly collinear particles originating from showering and hadronization of highly-energetic partons, and therefore can serve as their proxies. 
These partons are predominantly produced at the beginning of a heavy-ion (A+A) collision and may subsequently interact with the evolving medium generated in the collision. 
Specifically, in collisions with sufficiently high event activity (EA)---usually quantified by particle multiplicity or energy deposition within limited phase space---a quark-gluon plasma (QGP) is expected to form and in turn quench jets through collisional and radiative energy losses \cite{Miller:2007ri}. The resulting suppression of jet production rate in high-EA A+A collisions compared with that in \pp collisions was a strong early indication of QGP formation \cite{Adams:2003im} and continues to be a principle means for studying QGP physics \cite{STAR:2005gfr,ALICE:2010yje,CMS:2012aa}.
Jet quenching can be quantified by the nuclear modification factor, $R_\mathrm{A+A}^\mathrm{jet}$: the ratio of the jet yield per pair of colliding nucleons in A+A collisions to that in \pp collisions.
Traditionally, jet measurements in collisions of a heavy ion with a proton or deuteron ($p$ or $d$+A), usually referred to as small system collisions,
provide necessary references for  $R_\mathrm{A+A}^\mathrm{jet}$ by benchmarking the so-called cold nuclear matter (CNM) effects, i.e., effects related to the presence of a heavy nucleus in the collision but not to the creation of a QGP \cite{STAR:2007poe,STAR:2006xud,Adams:2003im}.

In this formulation, small system collisions are considered to be qualitatively different from high-EA A+A collisions and can be used as QGP-free benchmarks.  However this clear qualitative difference has become more nuanced with the observation of long-range correlations in soft particle production in high-multiplicity \pp collisions \cite{CMS:2010ifv}; such correlations are conventionally interpreted as resulting from QGP flow in A+A collisions. That first observation in \pp has motivated a broad and ongoing interest in studying small systems for QGP-like signals. The results have been fruitful, e.g., most flow-like signals have been observed in small system collisions, but have also raised new questions and motivated the search for evidence of jet quenching \cite{Nagle:2018nvi,Stankus:2016usz,Loizides:2016tew,Ohlson:2017jkm,McGlinchey:2016ssj}. Concurrent theoretical calculations on possible formation of small-volume QGP allow for results ranging from modest \cite{Tywoniuk:2014hta} to  significant \cite{Zakharov:2013gya} jet quenching, and therefore motivate the need for experimental exploration. 

Several measurements of $R_{p+\mathrm{Pb}}^\mathrm{jet}$ have been reported from $\sqrt{s_\mathrm{NN}}=\SI{5.02}{TeV}$ $p$+Pb collisions, where $\sqrt{s_\mathrm{NN}}$ is the center-of-mass energy per nucleon-nucleon pair. When not selected for EA, the $R_{p+\mathrm{Pb}}^\mathrm{jet}$ measurement is consistent with unity, i.e., no sign of jet suppression is observed \cite{ATLAS:2014cpa,Khachatryan:2016xdg,Adam:2015hoa,Adare:2015gla}.
However, when selecting events by EA at the ATLAS experiment, the $R_{p+\mathrm{Pb}}^\mathrm{jet}$ for high transverse momentum (\pT) jets is suppressed (enhanced) in high- (low-) EA events \cite{ATLAS:2014cpa}.  Tellingly, for central to $p$-going rapidities, the spectrum modification can be parametrized by the total jet energy ($E_\mathrm{jet}=\pT\cosh(y_\mathrm{jet})$, where $y_\mathrm{jet}$ is the jet rapidity).  This is intriguing because jets are proxies for hard scattered partons; therefore, the jet energy scaled by $2/\sqrt{s_\mathrm{NN}}$ can be used as the experimental approximation of $x_\mathrm{T}$ of the scattered parton in the colliding proton, which is related to the parton momentum fraction \xp. This interpretation is reinforced by a more recent measurement using $\sqrt{s_\mathrm{NN}}=\SI{8.16}{TeV}$ $p$+Pb collisions, in which dijets are used to further constrain the experimental estimate of \xp, and demonstrate a similar correlation between EA and \RpPbjet \cite{ATLAS:2023zfx}. Proposed causes for an \xp to EA correlation include  energy conservation (e.g., as in the Angantyr implementation in \PythiaEight \cite{Bierlich:2018xfw}), fluctuation of the proton size \cite{Alvioli:2014eda,McGlinchey:2016ssj,Alvioli:2017wou}, and color transparency between successive nucleon-nucleon collisions \cite{Kordell:2016njg}. Such a correlation could bias centrality classification of small system collisions: high- (low-) \xp collisions would have lower (higher) EA and therefore could be misclassified as a more peripheral (central) events. This would subsequently change the geometric scaling used to calculate centrality selected values of \RpdA. It is of interest to note that measurements of $Z$ bosons have also indicated a qualitatively similar selection bias in centrality selection for low-EA, i.e., peripheral, Pb + Pb events \cite{CMS:2021kvd}.

%The presence of a (anti-) correlation between \xp and EA has implications for centrality measurements: high- (low-)\xp events would produce less (more) EA signal. In small system collisions,  If there is an additional anti-correlation between \xp and EA, there may be a significant centrality bias: high- (low-)\xp events would produce less (more) EA signal. This in turn would result in \RpdA suppression (enhancement). 

%The correlation between EA and \RpdA in small system collisions continues to be an active area of research. EA is usually measured by soft particle production, which generally scales with the number of nucleons participating in the collision ($N_\mathrm{part}$), while the jet production (parameterized in part by \xp) scales with the number of binary nucleon-nucleon collisions ($N_\mathrm{coll}$). In A+A measurements, EA provides a good determination of the collision geometry, and the dynamic range of the EA is sufficiently large that it can be considered independent of \xp, even if the EA actually does include small fluctuations correlated to \xp.  In $p$/$d$+A collisions, the EA signal is much smaller and has large relative fluctuations, weakening its correspondence to the collision geometry and the scaling with \Npart. 

Subsequent to the first observation of the EA dependence of \RpPbjet, two measurements in $p$+Pb collisions were made by the ALICE experiment for jets up to around $\xp\approx0.05$.  The first measured inclusive jets with a modification to avoid possible centrality classification bias \cite{ALICE:2016faw}. The second measurement was of jet spectra per high-\pT hadron trigger (``h+jet'') \cite{ALICE:2017svf}, also referred to as a ``semi-inclusive'' measurement \cite{STAR:2017hhs},  not to be confused with the term in semi-inclusive deep-inelastic scattering. (Note: this current publication uses the convention as in the ALICE measurement.) Because of the per-trigger scaling, the semi-inclusive jet spectra may be compared directly between different EA classes without applying any potentially biased geometric scaling. Neither measurement found any dependence of jet spectra on EA. The semi-inclusive measurement also reported a limit on out-of-cone energy transport which, if it is the only mechanism responsible for jet yield suppression in $p$+A collisions, is inconsistent with the EA dependence of \RpPb at high \xp measured by the ATLAS experiment \cite{ALICE:2017svf,ATLAS:2014cpa}.

%In contrast with the lack of evidence for jet quenching in small system collisions in the above measurements
On the other hand, a recent publication reports the suppression of the production of high-\pT $\pi^{0}$s relative to that of direct photons in the 5\% highest EA $d$+Au collisions at \sNNcc measured with the PHENIX detector \cite{PHENIX:2023dxl}. The observed suppression is on the order of 20\% and, as $\pi^{0}$s may interact with a QGP while photons do not, may indicate a final-state effect.

This publication complements those results by reporting the first semi-inclusive jet measurements reaching high-\xp  (up to $\xp\approx0.4$) at the BNL Relativistic Heavy Ion Collider using $p$+Au collisions at \sNNcc.  We also compare dijet \pT imbalance and azimuthal distributions in high- and low-EA events to search further for possible jet quenching signals.%investigate if the dependence results from actual jet quenching instead of some other mechanism, such as those listed above. 

%%%%%%%%%%%%%%%%%%%%%%%%%%%%%%%%%%%%%%%
% Section: Experiment                 %
%%%%%%%%%%%%%%%%%%%%%%%%%%%%%%%%%%%%%%%
\section{Experiment, Data, and Methodology}\label{sec:Experiment}

\subsection{Event Selection}

RHIC provided $p$+Au collisions at $\sqrt{s_\mathrm{NN}}=\SI{200}{GeV}$ in 2015.
Minimum bias (MB) triggered events were collected by the STAR experiment based on the coincident signal in the Vertex Position Detectors (VPDs) \cite{Reed:2010zza}, located at both forward and backward pseudorapidities $|\eta_\mathrm{VPD}|\in[4.2,5.1]$. Additionally, events with high-momentum-transfer (high-$Q^{2}$) scatterings were selected by an online requirement using the  Barrel Electromagnetic Calorimeter (BEMC) \cite{STAR:2002ymp}, which has full azimuthal coverage  at midrapidity, $|\eta_\mathrm{BEMC}|\le1$.
Specifically, the online trigger selection requires at least one BEMC tower with transverse energy of at least \SI{2.5}{GeV}.
These events are designated in this publication as ``high tower'' (HT) events, and the leading tower energy as ``\ETtrig''. To select high jet \pT events desired for this study, the threshold for \ETtrig is raised to \SI{4}{GeV} as an offline event selection. No additional criteria are added to discriminate between possible signal sources, whether photon, hadron, or otherwise.

STAR has two Zero Degree Calorimeters (ZDCs)  located \SI{18}{m} away on either side of the nominal interaction point \cite{Xu:2016alq}. ZDCs measure neutrons not participating directly in the collisions and the ZDC coincidence rate (ZDCx) measures the instantaneous luminosity accompanying the triggered collisions. Since the rate of pileup events, which occur slightly before or after the collisions of interest, increase with luminosity, this work only analyzes events with ZDCx less than \SI{20}{kHz}.
%This \SI{20}{kHz} maximum ZDCx value corresponds to a \pAu collision rate of approximately \SI{930}{kHz}, or 36 collisions per the maximum TPC drift time of \SI{38.5}{\mu{}s}.
This \SI{20}{kHz} maximum ZDCx value corresponds to a \pAu collision rate of approximately \SI{930}{kHz}, meaning on average 36 collisions occur during the \SI{38.5}{\mu{}s} it takes for signal charge to drift from the TPC’s central membrane to the readout plane. A lower bound of \SI{5}{kHz} is also applied on ZDCx due to limited statistics below \SI{5}{kHz}.

Event vertex locations are determined from the projection of charged particle tracks in STAR's Time Projection Chamber (TPC) gas volume \cite{Anderson:2003ur}.
Events with $V_\mathrm{z}$, the vertex position along the $z$-axis parallel to the beam line, within \SI{10}{cm} of the center of the STAR detector are selected.  Of those, any event with a $V_\mathrm{z}$ greater than \SI{6}{cm} away from the vertex $z$-position measured by the VPD ($V_\mathrm{z,VPD}$) is rejected. This is because the VPD is a fast detector, and therefore resilient to pileup collisions.

Using the above online and offline selections, this publication reports on measurements from 3.7 million MB and 135 million HT collisions.

The event selections are summarized in Table~\ref{tab:event_selection}.

\begin{table}[H]
\centering
\caption{Event Selections} \label{tab:event_selection}
\begin{tabular}{@{}ll@{}}
\toprule
\multicolumn{2}{c}{\textit{Online Event Selection}}\\ %\cmidrule{1-1}
Minimum Bias (MB): &  Signals in both VPDs \\
\addlinespace[.3em]
High Tower (HT): &  Signals in both VPDs \\
&  High energy BEMC tower $>\SI{2.5}{GeV}$ \\
% \end{tabular}
% \begin{tabuler}
& $|V_\mathrm{z,VPD}|<\SI{30}{cm}$ \\
% \multicolumn{2}{l}{Signal in East and West Vertex Position Detector (VPD)} \\
\addlinespace
\multicolumn{2}{c}{\textit{Offline Event Selection}}\\
Reconstructed vertex: &  $|V_\mathrm{z}|<\SI{10}{cm}$ \\
                      &  $|V_\mathrm{z}-V_\mathrm{z,VPD}|<\SI{6}{cm}$\\
\addlinespace[.3em]
Instantaneous luminosity: &  $\mathrm{ZDCx}\in [5,20]\;\mathrm{kHz}$ \\
\addlinespace[.3em]
Leading BEMC tower (HT): &  $E_\mathrm{T}^\mathrm{trig} > \SI{4}{GeV}$ \\
 \bottomrule
% \multicolumn{2}{l}{\footnotesize$^*$Lower $\rm ZDCx$ bound due to limited statistics below $\SI{5}{kHz}$}\\
% \multicolumn{2}{l}{\footnotesize$^\dag$ HT events are subdivided in Fig.~\ref{fig:UE_vs_EAbbc} [4,8), [8,12) and [12,30)} \\ 
 %Except for graduated ranges listed in Fig.~\ref{fig:UE_vs_EAbbc}} \\
\end{tabular}
\end{table}

\subsection{Event Activity}\label{sec:EA}

STAR traditionally uses charged particle multiplicity at midrapidity to classify EA or centrality for Au + Au collisions \cite{Miller:2007ri}. However, in small system collisions, particle production from hard scatterings contribute a significant fraction to the total charged particle multiplicity, resulting in a localized autocorrelation between jet production and EA. A possible solution is to separate the acceptance of the EA measurement from that of the jets. One way to do this is to measure the underlying event (UE) ``beneath'' the hard scattering, i.e., charged particle density ($\frac{\mathrm{d}N^{2}_\mathrm{ch}}{\mathrm{d}\eta d\phi}$) transverse to the jet axis. Another way is to classify EA with particle production at large rapidities, well separated from midrapidity jets. The latter is the method employed in this publication by using STAR's Beam-Beam Counters (BBCs).

The BBCs are located $\pm\SI{3.75}{m}$ from STAR's nominal interaction point \cite{Whitten:2008zz}. They each consist of annuli of hexagonal-shaped scintillating tiles with full azimuthal coverage in aggregate, receiving signals from the charged particle flux.  This work utilizes the inner annuli of the BBC in the Au-going direction ($\eta_\mathrm{BBC}\in[-5,-3.4]$), and quantifies EA as the sum of the ADC values from the response of those tiles (``\EAbbc''). This provides a wide rapidity gap between the EA acceptance and the jets measured at midrapidity ($|\eta|\le0.6$). The rapidity gap is large enough that at RHIC kinematics, when one jet is found at midrapidity, the recoiling jet of a dijet pair cannot reach the EA acceptance  \cite{STAR:2018yxi}. BBC signals in the Au-going direction are preferred to those in the $p$-going direction because they have a greater range of measured EA.

\begin{figure}[t!]
    \centering
    \includegraphics[width=0.45\textwidth, trim=0.3cm 1.0cm 0.0cm 0.0cm, clip]
    {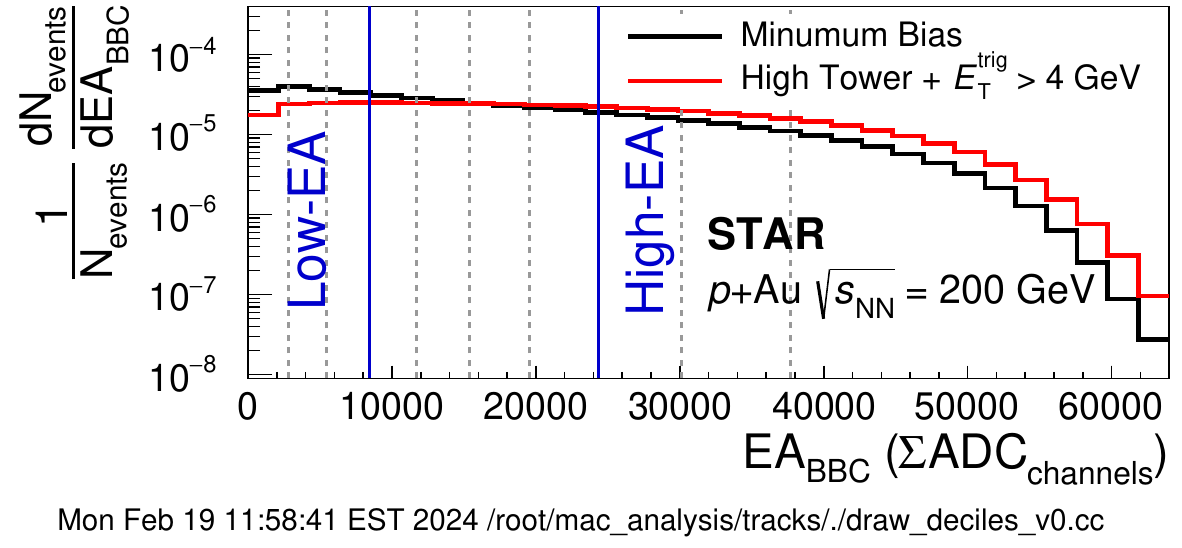}
    \caption{Distributions of $\mathrm{EA}_\mathrm{BBC}$ from MB and HT events with the latter satisfying an offline trigger of $E_\mathrm{T}^\mathrm{trig}>\SI{4}{GeV}$. Vertical dashed lines indicate 10\% deciles of the MB distribution. Low- and high-EA are defined as the lowest and highest 30\% of the MB distribution, and are indicated by the vertical solid lines and labels.}
    \label{fig:EA_dist}
\end{figure}

The distributions of EA for MB and HT events are given in Fig.~\ref{fig:EA_dist}.  An event is classified as low- (high-) EA if its EA signal is smaller (larger) than the 30\% (70\%) decile of the EA distribution in MB events (as indicated by vertical solid lines in Fig.~\ref{fig:EA_dist}).  HT events skew the EA toward higher values, such that about 19\% (42\%) of HT collisions are classified as low- (high-)EA.
This is consistent with the correlation observed in A+A collisions: higher EA should correlate with more central events, which have more binary nucleon-nucleon collisions, and therefore higher probability of high-$Q^2$ parton interactions.

In the recorded \pAu sample, signals of tiles in the BBC, which was originally designed as a trigger detector \cite{Judd:2018zbg}, were frequently saturated so that \EAbbc isn't directly proportional to the number of minimum ionizing particles (MIPs) in the BBC acceptance. Therefore, we justify using the BBC as an EA classifier by measuring correlations between \EAbbc and the charged particle density at midrapidity, as shown in Figs.~\ref{fig:EA_vs_EA_MB} and \ref{fig:EA_vs_EA_HT} in the appendix.  The UE charged particle density is measured using charged particle tracks in the TPC, as introduced in Sec.~\ref{sec:jets_track_towers}. In MB events, the charged particle density using the full azimuthal coverage of the TPC is reported. In HT events, to avoid the localized autocorrelation introduced by the charged particles in the triggering dijet pair, only the charged particle density azimuthally transverse to the trigger ($|\phi_\mathrm{ch}-\phi_\mathrm{trigger}|\in[\pi/3,2\pi/3]$) is measured. While the correlation between EA and the UE charged particle density is broad, it is on average positive and monotonic.  Qualitatively, this correlation is similar to that observed for EAs measured at $\eta\in[-3.8,-2.4]$ and using the ZDC \cite{Xu:2016alq} in $d$+Au events \cite{STAR:2014qsy}. 

The fiducial acceptances of the measurements presented in this paper are summarized in Table~\ref{tab:rapididy_selection}.

\begin{table}[H]
\centering
\caption{Measurement Fiducial Acceptances} \label{tab:rapididy_selection}
\begin{tabular}{@{}ll@{}}
\toprule
Measurement & Acceptance \\ \midrule
\multicolumn{2}{c}{\textit{Pseudorapidities$^*$}} \\
$\mathrm{EA}_\mathrm{BBC}$ & $\eta\in[-3.4, -5]$     \\
UE charged particle density& $|\eta|<0.9$ \\
TPC tracks & $|\eta|<1.0$ \\
BEMC towers & $|\eta|<1.0$ \\
Jets & $|\eta|<0.6$ \\ 
\addlinespace
\multicolumn{2}{c}{\textit{UE charged particle density azimuths}} \\
MB events$^\dag$ & $\phi_\mathrm{ch}\in[0,2\pi]$ \\
HT events$^\dag$ & $|\phi_\mathrm{ch}-\phi_\mathrm{trigger}|\in[\pi/3,2\pi/3]$ \\
HT events with jets${^\ddag}$ & $|\phi_\mathrm{ch}-\phi^\mathrm{lead}_\mathrm{jet}|\in[\pi/3,2\pi/3]$ \\
\bottomrule
\multicolumn{2}{l}{\footnotesize $^*$For details on tracks, towers, and jets see Sec.~\ref{sec:jets_track_towers}} \\
\multicolumn{2}{l}{\footnotesize $^\dag$See Fig.~\ref{fig:UE_vs_EAbbc}~~~$^\ddag$See Fig.~\ref{fig:UEplot}} \\
\end{tabular}
\end{table}

\subsection{Jets: Tracks and Towers}\label{sec:jets_track_towers}

The TPC measures charged particle tracks and their kinematics via ionization in its gas volume. The TPC has full azimuthal coverage within $|\eta|\lesssim1.3$. This analysis only uses tracks out to $|\eta|<1.0$ in order to match the BEMC acceptance. Only tracks with $\pT\in[0.2,30]\;\mathrm{GeV}/c$ are analyzed, in order to avoid poor detection efficiency at low \pT and poor resolution at high \pT. All selected tracks are also required to have at least 20 TPC hits for their reconstruction, pass within \SI{1}{cm} of  the primary vertex, and contain at least 52\% of possible TPC hits along their trajectories.

The BEMC provides energy measurements for electromagnetically interacting particles within $|\eta|<1.0$ \cite{STAR:2002ymp}. It overlaps the fiducial coverage of the TPC and has an angular granularity of about 0.05 in both $\phi$ and $\eta$. The transverse energy (\ET) component of the calorimeter energy is determined from the tower location relative to the collision vertex via an inverse factor of $\cosh{(\eta_\mathrm{tower})}$. The tower \ET limits match the track \pT limits of $[0.2,30]\;\mathrm{GeV}$. Of the 4800 calorimeter towers, 318 towers are excluded due to malfunction during data taking.

In each event, tracks and towers are clustered into jets using the \antikT algorithm \cite{Cacciari_2008} with a resolution parameter $\Rjet=0.4$. To limit boundary effects, only jets within $|\eta_\mathrm{jet}|<1-R_\mathrm{jet}$ are used. If a charged particle's trajectory extrapolates to a tower, it is assumed that the track and the corresponding calorimeter energy (if any) result from the same particle. To avoid double counting that particle's energy, its value is subtracted from the tower energy. Here the particle energy is calculated based on its momentum measured in the TPC and the assumption that it is a pion. Any tower with $\ET<\SI{0.2}{GeV}$ after subtraction is discarded. While this procedure, referred to as 100\% hadronic correction, could potentially over-correct any double-counting, it results in better jet energy resolution and the potential over-correction is accounted for during unfolding \cite{STAR:2014wox}. 

The track and BEMC selections are summarized in Table~\ref{tab:track_selection}.

\begin{table}[H]
\centering
\caption{Track and BEMC Tower Selection Criteria}\label{tab:track_selection}
\begin{tabular}{@{}ll@{}}
\toprule
Selection & Value \\ \midrule
\multicolumn{2}{c}{\textit{Track Cuts}} \\
Number of TPC fit points & $\ge20$     \\
Ratio of fit points to possible points & $\ge0.52$ \\
Distance of closest approach to vertex & $\le\SI{1}{cm}$ \\
Range of $p_\mathrm{T}$ & $[0.2,30]\;\mathrm{GeV}/c$ \\
\multicolumn{2}{c}{\textit{BEMC Cuts}} \\
Range of $\ET^*$ & $[0.2,30]\;\mathrm{GeV}$ \\
\bottomrule
\multicolumn{2}{l}{~~\footnotesize*
\begin{minipage}[t]{18em}The \SI{0.2}{GeV} cut is applied after the 100\% hadronic correction.\end{minipage}}
\end{tabular}
\end{table}

\subsection{Track and Jet Corrections}\label{sec:tr_jet_corrections}

The TPC tracking efficiency is evaluated by simulating the STAR detector response to $p$, $\bar{p}$, $\pi^{\pm}$, and $K^{\pm}$ via \GeantThree \cite{Brun:1994aa}. These simulated signals are embedded into HT events to reflect realistic running conditions, and then reconstructed using the standard STAR framework. The embedded spectra are weighted according to measured particle yields in $\sqrt{s_\mathrm{NN}} = \SI{200}{GeV}$  $d$+Au collisions \cite{STAR:2006nmo,STAR:2006xud} to obtain the tracking efficiency for the expected mixture of charged particles. It is worth noting that spectral $K$ abundances in $d$+Au collisions have not been measured at STAR; therefore appropriately scaled $\pi^{\pm}$ spectra are used in their place \cite{STAR:2006xud}. 

Since the tracking efficiency begins to decrease when approaching the boundary of the TPC acceptance, the measurement of charged particle density is limited to $|\eta|\le0.9$. Within that range, in events with $\mathrm{ZDCx}\in[5,8]\;\mathrm{kHz}$ the track reconstruction efficiency is 70\% for a track $p_\mathrm{T}$ at $0.2\;\mathrm{GeV}/c$ and rises to an approximately constant value of 85\% beyond $0.6\;\mathrm{GeV}/c$, resulting in an average efficiency of 76\%. The efficiency then drops with increasing luminosity to an average value of 66\% when $\mathrm{ZDCx}\in[17,20]\;\mathrm{kHz}$.
When the luminosity dependent decrease in tracking efficiency is accounted for, the average number of tracks in the TPC increases by about 0.5 tracks in high luminosity events relative to low luminosity events. This increase in track density as a function of luminosity is attributed to the presence of residual pileup tracks which are subtracted from the final reported charged particle densities.

Detector effects on jets are corrected using a similar embedding methodology. Dijet events, simulated with \PythiaSix \cite{Sjostrand:2006za}, are propagated through the \GeantThree simulation of the STAR detector before being embedded into MB events, from which tracks and towers are reconstructed and clustered into detector-level (``raw'') jets. Only events with a reconstructed leading-tower, i.e., the trigger tower, with $\ET\ge\SI{4}{GeV}$ are used. It is found in the simulations that the HT trigger efficiency is independent of EA at all jet momenta. Therefore, even though the trigger efficiency is not explicitly corrected for in this publication, its effect is expected to cancel in the ratios of jet quantities between high- and low-EA events. 

Truth-level jets are clustered directly from the simulated \PythiaSix events. The truth- and detector-level jets are matched geometrically in $\phi$ and $\eta$ by pairing jets with $\sqrt{(\Delta\phi)^2+(\Delta\eta)^2}\le0.4$, preferentially matching the highest-\pT jets if multiple candidates are found. The jet energy scale and resolution (JES and JER) are defined as the average difference in \pT between matched truth- and detector-level jets, and the standard deviation of that \pT difference distribution, respectively. The JES($\pm$JER) values range from $-1.1(\pm2.1)\,\mathrm{GeV}/c$ for \SI{15}{GeV/\mathit{c}} jets, to $-4.1(\pm4.5)\,\mathrm{GeV}/c$ for \SI{40}{GeV/\mathit{c}} jets. While JES and JER are not used directly in any correction, they can quantify the detector performance. It is worth noting that the JES is comparable for high- and low-EA events.

The measured jet spectra are corrected using the Bayesian unfolding procedure implemented in the RooUnfold package \cite{Adye:2011gm} with six iterations and \PythiaSix distributions as the priors. The unfolding procedure uses a response matrix filled with matched truth- and detector-level jets to correct for the JES and JER.
The additional effects of fake detector-level jets and missed truth-level jets are corrected for using the $p_\mathrm{T,jet}$-dependent relative abundances of the unmatched jets remaining from the matching procedure.

\subsection{Systematic Uncertainties for Tracks, Jets, and Towers}

Systematic uncertainties in charged particle density measurements include the following: 5\% tracking efficiency uncertainty from imperfect detector simulation; an additional 2\% tracking efficiency uncertainty accounting for correlations between luminosity and EA; a subpercent uncertainty for the use of particle spectra in $d$+Au (as opposed to $p$+Au) collisions to weight embedding samples; a subpercent uncertainty for the unfolding method (bin-by-bin vs 2D Bayesian unfolding); and pileup subtraction uncertainties. The pileup uncertainty, arising mostly from parametrizing the increase of the charged particle density as a function of ZDCx after efficiency correction, ranges from 6\%-10\% and dominates other sources. When adding all individual sources in quadrature, the overall charged particle density uncertainty is 7\%-11\%.
The EA and trigger tower \ET values are detector-level characterizations of each event, and are uncorrected quantities without any assigned systematic uncertainties.

Jets are corrected for detector efficiency, acceptance, and resolution with the unfolding procedure as discussed in Sec.~\ref{sec:tr_jet_corrections}. Systematic uncertainties are evaluated by individually varying the response matrices, miss rates, and fake rates, and repeating the entire correction procedure to estimate the magnitude of the uncertainty. The tracking efficiency uncertainties are assessed by randomly removing reconstructed tracks according to the tracking uncertainty values. The tower energy scale uncertainty (3.8\%) is applied by uniformly augmenting all tower energies. Collectively these uncertainties constitute 15\%-20\% when added in quadrature,  and are the dominant components of the jet spectrum systematic uncertainties. Varying the hadronic correction from 100\%, as discussed in Section~\ref{sec:jets_track_towers}, to 50\% contributes about 2\%-4\% relative uncertainty. The stability of the iterative Bayesian unfolding is quantified by the average difference in the unfolded results when the number of iterations used is varied by $\pm2$ from the nominal value of 6. The shape of the prior distribution is reweighted to a distribution in \texttt{HERWIG} (from the default \PythiaSix distribution), resulting in an uncertainty of about 10\% in the first jet \pT bin and from 0\%-5\% in the remaining bins.  Finally, over-estimation of the background by embedding \PythiaSix events into MB $p$ + Au data contributes 1.6-2.8\% uncertainty.

With the exception of the contribution from the additional 2\% tracking efficiency uncertainty, the jet spectrum systematic uncertainties mostly cancel in the high-EA to low-EA ratios. The largest remaining uncertainties, resulting from the hadronic correction and embedding over-estimation, are between 1\%-2\%. The resulting overall systematic uncertainties of the jet spectrum ratios are on the order of 3\%-9\%.

\subsection{Semi-Inclusive Jet Spectra}

The semi-inclusive jet spectra are the $p_\mathrm{T}$ spectra of jets in events with an offline trigger (i.e., a BEMC tower with $\ET\ge\SI{4}{GeV}$), normalized by the number of these triggered events:
 $\frac{1}{N_\mathrm{trig}} \frac{\mathrm{d}N_\mathrm{jet}^{3}}{\mathrm{d}p_\mathrm{T}\mathrm{d}\eta\mathrm{d}\phi}$. Four different spectra are reported in Fig.~\ref{fig:semi_inc_jets}: near- and recoil-side spectra in high- and low-EA events. Near- and recoil-side spectra are defined as jets within $|\phi_\mathrm{jet}-\phi_\mathrm{trig}|<\pi/3$ and  $|\phi_\mathrm{jet}-\phi_\mathrm{trig}|>2\pi/3$, respectively. 
 Note that most triggered events have no jets within the reported \pT range, thus the integrals of the jet spectra are less than one. Additionally, the ratios of semi-inclusive spectra in high- and low-EA events are reported.

\subsection{Dijet Azimuthal Separation ($|\Delta\phi|$) and Momentum Imbalance (\AJ)}\label{sec:method_Aphi}

% Both \pT balance and dijet azimuthal differences are proposed methods of probing for QGP in A+A collisions \cite{STAR:2016dfv,CMS:2011iwn,ATLAS:2010isq,deramoMoliereScatteringQuarkgluon2019,Mueller:2016gko}, and both are, to first order, independent of an EA-to-\xp correlation.

The modification of dijet \pT imbalance and relative azimuthal distribution have been proposed as methods to probe for jet quenching by the QGP in A+A collisions \cite{STAR:2016dfv,CMS:2011iwn,ATLAS:2010isq,deramoMoliereScatteringQuarkgluon2019,Mueller:2016gko}. These are of particular interest because, to first order, the shapes of their distributions are independent of any EA-to-\xp correlation. This is not true for the semi-inclusive analysis, where an EA-to-\xp correlation would be manifested in the semi-inclusive jet spectra normalization.
The distribution of the azimuthal separation between the leading and subleading jets 
($|\Delta\phi|\equiv|\phi_\mathrm{jet}^\mathrm{lead}-\phi_\mathrm{jet}^\mathrm{sub}|$) 
is reported for all dijets with $|\Delta\phi|>\pi/2$.  Additionally, the distribution of the $p_\mathrm{T,jet}^\mathrm{raw}$ imbalance  $A_\mathrm{J}\equiv\frac{p_\mathrm{T,jet}^\mathrm{raw,lead}-p_\mathrm{T,jet}^\mathrm{raw,sub}}{p_\mathrm{T,jet}^\mathrm{raw,lead}+p_\mathrm{T,jet}^\mathrm{raw,sub}}$ is reported for  all dijet pairs with $\absDphi>(\pi-0.4)$, where \pTrawlead and \pTrawsub are the detector-level \pT for leading and subleading jets of the dijet pair, respectively. All these distributions are normalized to unity for shape comparisons.

Because the JES and JER distributions are not discernibly different between high- and low-EA events, detector effects cancel when detector-level distributions are taken in ratio. This justifies the presentation of the detector-level dijet momentum imbalance and azimuthal separation ratios. However, the dijet ratios could also be biased from anything that affects the selection cuts for $p_\mathrm{T,jet}^\mathrm{raw,lead}$ and $p_\mathrm{T,jet}^\mathrm{raw,sub}$, even if they do not significantly affect the JES and JER. Two possible such differences exist between the high-EA and low-EA events and are accounted for. First, high-EA events have modest increases in underlying event and pileup activity. Second, high-EA events have a higher mean ZDCx value which introduces a modest change in the tracking efficiency.  To account for the first effect, the high- and low-EA collision events are divided into groups selected by ZDCx. Within each ZDCx selection, the average number $(N_\mathrm{excess})$ and \pT spectrum ($\mathrm{dN}/\mathrm{d}p_\mathrm{T,raw}^\mathrm{excess}$) of particles in excess in high-EA events are measured. Then each low-EA event is augmented with particles, whose multiplicity is drawn from a Poisson distribution with mean equal to $N_\mathrm{excess}$ and transverse momenta sampled from $\mathrm{dN}/\mathrm{d}p_\mathrm{T,raw}^\mathrm{excess}$. Additionally, high-EA events are re-weighted to match the ZDCx distribution of the low-EA events. 
With these modifications, the high-EA and low-EA events are clustered, dijets are selected, and their detector-level distributions and ratios are reported.

\subsection{Experimental Conditions Inherent in Results}

It is worth noting several experimental conditions which are inherent in the reported results and should be kept in mind in any attempt to compare with theory. First, the EA definition is empirically justified, and therefore direct connection to actual particle densities in the BBC acceptance is not available. Additionally, the MB trigger and the \SI{4}{GeV} tower trigger are not corrected for trigger biases.

%%%%%%%%%%%%%%%%%%%%%%%%%%%%%%%%%%%%%%%
% Section: Measurements               %
%%%%%%%%%%%%%%%%%%%%%%%%%%%%%%%%%%%%%%%
\section{Results}\label{sec:Results}

\subsection{Underlying Event}% Note: pp data at HEPdata at https://www.hepdata.net/record/ins1771348

The fully-corrected UE charged particle densities are plotted in Fig.~\ref{fig:UE_vs_EAbbc} for ten ranges of EA. The EA ranges are selected such that each contains a decile of the \EAbbc distribution in MB events. The HT-triggered events are separated by their values of \ETtrig into three groups.

The charged particle densities increase monotonically with increasing EA for both MB and HT events. The lowest values are consistent with the underlying event activity measured in $\sqrt{s}=\SI{200}{GeV}$ \pp collisions which contain low-\pT ($5\!-\!7\,\mathrm{GeV}/c$) charged jets~\cite{STAR:2019cie}. Herein, low-EA $p$+Au collisions appear to be similar to $p\!+\!p$ collisions. In the highest EA events, the charged particle densities are around a factor of two higher than that in \pp events.
The figure shows that UE charged particle density scales with EA, and also hints that the underlying event may be anti-correlated with increasing trigger energy. %These are qualitatively consistent with the underlying event measurements in \pp collisions~\cite{STAR:2019cie}. 

\subsection{Event Activity vs. Leading Jet \pT}\label{sec:eabias}

\begin{figure}[t!]
    \centering
    \includegraphics[width=0.48\textwidth, trim=0.0cm 0.0cm 0.0cm 0.0cm, clip]
    {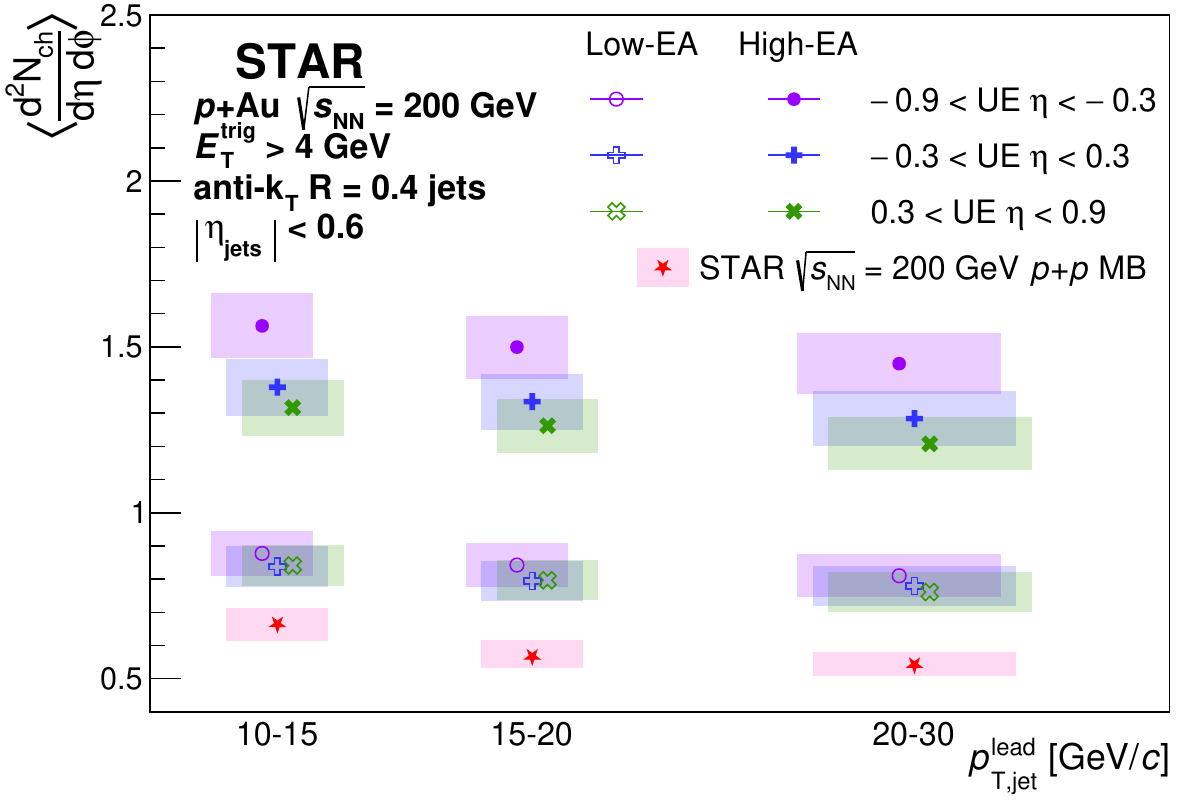}
    \caption{Charged particle density transverse to the leading jet ($|\phi_\mathrm{jet}^\mathrm{lead}-\phi_\mathrm{ch}|\in[\pi/3,2\pi/3]$) as a function of EA, UE $\eta$, and $p_\mathrm{T,jet}^\mathrm{lead}$ for HT ($E_\mathrm{T}^\mathrm{trig}>\SI{4}{GeV}$) events with $p_\mathrm{T,jet}^\mathrm{raw,lead}>4~\mathrm{GeV}/c$. Solid (open) markers represent high-EA (low-EA) events with their highest-$p_\mathrm{T}$ jet as specified along the x-axis. The circles, +'s, and X's represent the multiplicity densities in the Au-, mid-, and $p$-going pseudorapidities, respectively. Statistical errors are smaller than the plotted symbols. The markers in each $p_\mathrm{T,jet}^\mathrm{lead}$ selection have small horizontal offsets for visual clarity. The red stars show the values for STAR UE in \pp collisions for $p_\mathrm{T,jet}^\mathrm{lead}$ $\in$ (11,15), (15,20), (20,25)~GeV/$c$~\cite{STAR:2019cie}. The shaded boxes show the size of the systematic uncertainties}
    \label{fig:UEplot}
\end{figure}

Figure~\ref{fig:UEplot} reports the UE charged particle density differentially with respect to three variables:

\begin{enumerate}
    \item Leading jet \pT: The trigger tower is required to  be either within the leading jet or azimuthally recoiling from it ($|\phi_\mathrm{tower}-\phi_\mathrm{jet}^\mathrm{lead}|>(\pi-R_\mathrm{jet})$). Data are shown for events with $p_\mathrm{T,jet}^\mathrm{raw,lead}>4~\mathrm{GeV}/c$ in three ranges of leading jet \pT: (10,15), (15,20), and (20,30) GeV/$c$.
    \item High-EA and low-EA, as defined in Sec.~\ref{sec:EA}.
    \item Pseudorapidity of charged particles transverse to the leading jet ($|\phi_\mathrm{ch}-\phi_\mathrm{jet}^\mathrm{lead}|\in[\pi/3,2\pi/3]$) in the TPC: %Au-going ($\eta\in(-0.9,-0.3)$), mid-rapidity ($\eta\in(-0.3,0.3)$), and $p$-going ($\eta\in(0.3,0.9)$) regions. 
    \begin{itemize}
        \item Au-going, $\eta\in(-0.9,-0.3)$
        \item midrapidity, $\eta\in(-0.3,0.3)$
        \item $p$-going, $\eta\in(0.3,0.9)$
    \end{itemize}
\end{enumerate}

\begin{figure}[t!]
    \centering
    \includegraphics[width=0.45\textwidth, trim=1.3cm 0.70cm 0.0cm 0.0cm, clip]
    {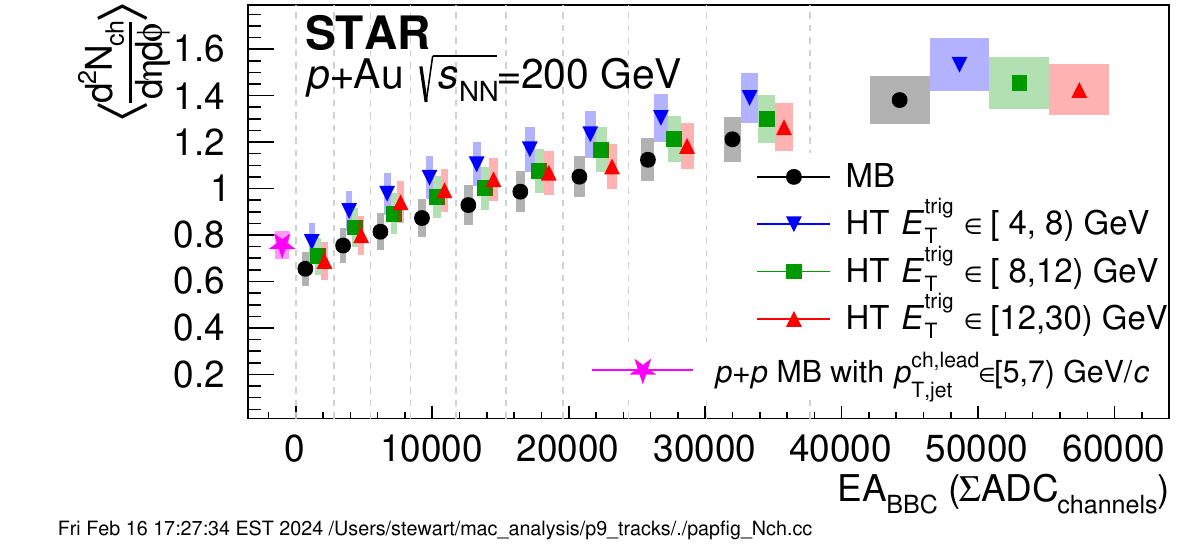}
    \caption{Density of charged particles with $\pT>0.2$~GeV/$c$ and $|\eta|\le0.9$ in ten ranges of EA for MB events and HT events selected with several offline trigger values.  Markers are horizontally offset within each range of EA for visual clarity. The star plotting symbol is for \SI{200}{GeV} $p\!+\!p$ collisions containing a low-$p_\mathrm{T}$ jet ~\cite{STAR:2019cie}, and is not associated with any specific EA value. The error bars are statistical, while the shaded boxes show the size of the systemmatic uncertainties.}
    \label{fig:UE_vs_EAbbc}
\end{figure}

As expected from the asymmetry of the colliding nuclei, the UE is higher at Au-going rapidities. Figure~\ref{fig:UEplot} also confirms the correlation between UE charged particle density and EA expected from Fig.~\ref{fig:UE_vs_EAbbc}: events selected at high-EA (low-EA) have correspondingly higher (lower) mean densities for all values of jet \pT. %When equating the trigger \ET cuts in Fig.~\ref{fig:UE_vs_EAbbc} and the lead jet \pT in \fig{UEplot}, a correlation in which UE and EA decrease with increasing the $Q^2$ of the hard scattering is again suggested.
As is the case in Fig.~\ref{fig:UE_vs_EAbbc}, Fig.~\ref{fig:UEplot} also suggests that UE decreases with increasing jet \pT.

\begin{figure}[t!] \centering
    \includegraphics[width=0.45\textwidth, clip]
    {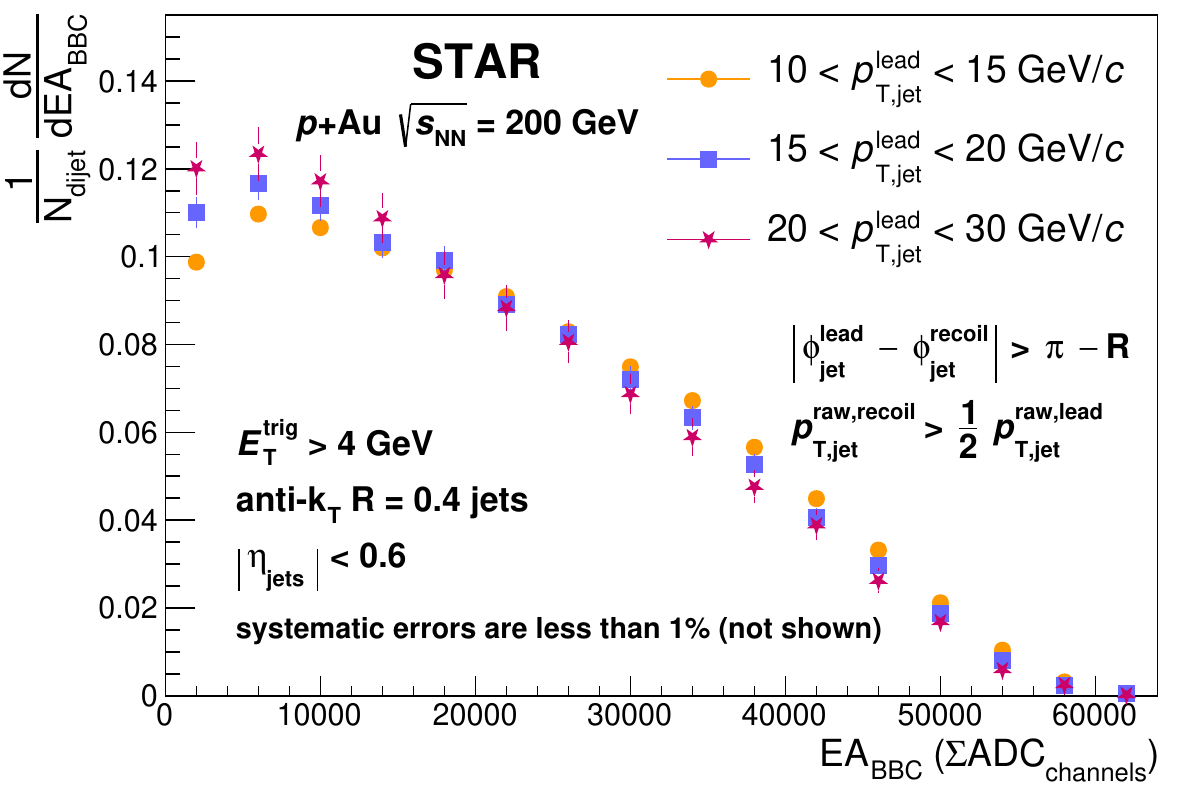}
    \caption{ %
        Normalized distributions of EA in HT ($E_\mathrm{T}^\mathrm{trig}>\SI{4}{GeV}$) events selected by three ranges of leading jet $p_\mathrm{T}$. Events require $p_\mathrm{T,jet}^\mathrm{raw,lead}>4~\mathrm{GeV}/c$ and the presence of a recoil jet.
    }\label{fig:EA_vs_jets}
\end{figure}

To explore the correlation between hard process and EA further, Fig.~\ref{fig:EA_vs_jets} presents the distributions of EA for three ranges of \pTlead, as used in \fig{UEplot}. As already noted, in \SI{200}{GeV} collisions when one member of a dijet pair is measured in the TPC ($|\eta|<0.6$), it is not kinematically accessible for the other jet to reach the BBC and therefore cause an anticorrelation.
 Nevertheless, to ensure this does not affect the measurement, the event selection for Fig.~\ref{fig:EA_vs_jets} requires both members of a dijet pair to be found fully within the TPC acceptance. This is done using the two highest-\pT detector-level jets (\pTrawlead and \pTrawrecoil) and requiring that: $\pTrawlead>\SI{4}{GeV/\textit{c}}$, $\pTrawrecoil>\frac{1}{2}\pTrawlead$, and $\left(\left|\phi_\mathrm{jet}^\mathrm{lead}-\phi_\mathrm{jet}^\mathrm{recoil}\right|>\pi-R_\mathrm{jet}\right)$.
Within these events,  there is a statistically significant drop of mean EA with increasing \pTlead: $21190\pm{}80$, $20300\pm{}100$ and $19500\pm{}200$, respectively, presenting a clear anticorrelation between EA measured in the BBC and the average momentum transfer of a hard scattering at midrapidity.
% 10 < p_{T,lead} < 15 GeV/#it{c}    21190 +- 80
% 15 < p_{T,lead} < 20 GeV/#it{c}    20300 +- 100
% 20 < p_{T,lead} < 30 GeV/#it{c}    19500 +- 200

\subsection{Semi-Inclusive Jet Spectra}\label{sec:S}

\begin{figure}[t!] \centering
    \includegraphics[width=0.47\textwidth, trim=0.8cm 1.0cm 0.9cm 0.0cm, clip]{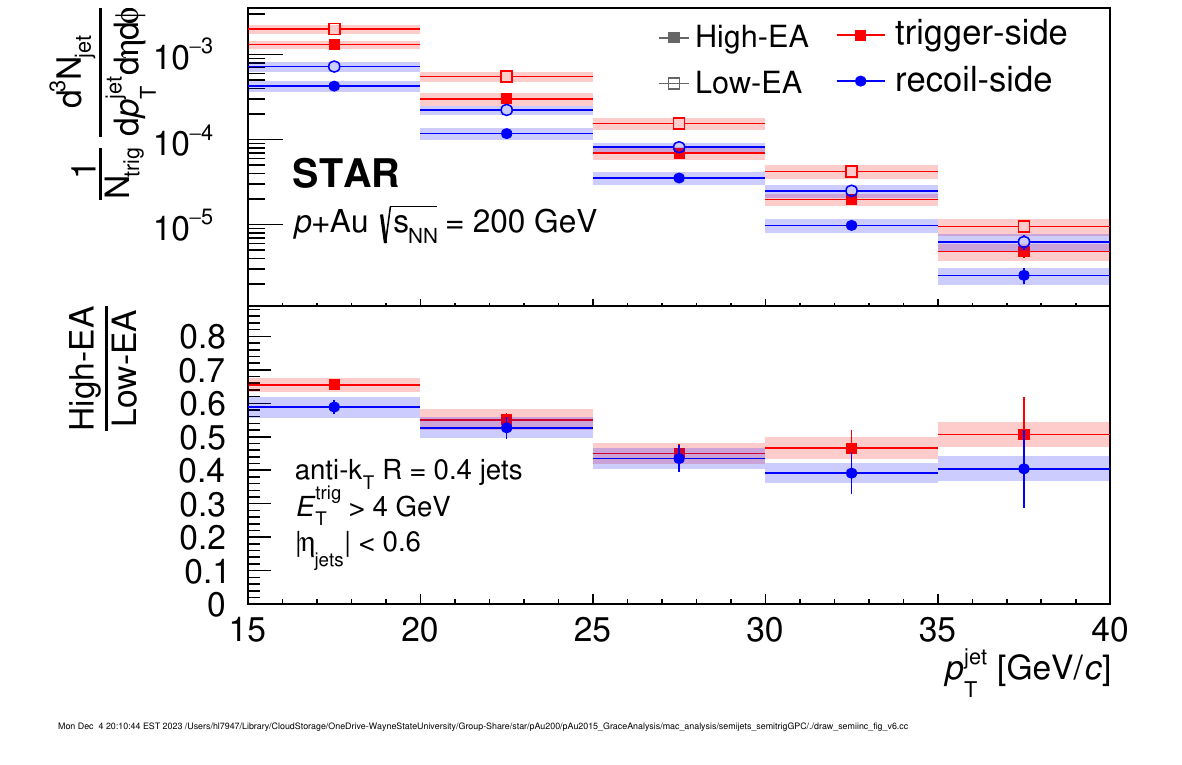}
    \caption{ Top panel: jet spectra per trigger for trigger-side ($|\phi_\mathrm{jet}-\phi_\mathrm{trig}|<\pi/3$) and recoil-side ($|\phi_\mathrm{jet}-\phi_\mathrm{trig}|>2\pi/3$). Jets are of $R=0.4$, and the offline trigger threshold is $\ETtrig>\SI{4}{GeV}$. Spectra are shown for high- and low-EA events. Bottom panel: ratio of the semi-inclusive jet spectra in high-EA to low-EA events. The error bars are statistical, while the shaded boxes show the size of the systematic uncertainties.}\label{fig:semi_inc_jets}
\end{figure}

Figure~\ref{fig:semi_inc_jets} reports the first fully-corrected semi-inclusive jet spectra in small system collisions at RHIC. The top panel is the jet spectra per trigger, counting only jets azimuthally within $\pi/3$ of the triggering tower for trigger-side jets, while recoil-side jets are required to be azimuthally within $\pi/3$ of the direction opposite the trigger tower. The spectrum of the recoiling jets has significantly fewer counts per trigger relative to that of the trigger-side jets due to a large fraction of recoiling jets falling outside of the detector's acceptance ($|\eta_\mathrm{jets}|<0.6$)~\cite{STAR:2018yxi}. As shown in the bottom panel, both spectra are distinctly suppressed in high-EA events relative to low-EA events. Notably, the suppression for both trigger- and recoil-side jets is comparable. This is qualitatively different than jet-spectrum suppression attributed to jet quenching in the QGP \cite{STAR:2012civ}; there, the QGP causes a selection bias and resulting path-length difference which suppresses the recoil-side jets more than the trigger-side jets.

If the anti-correlation between EA and jet \pT, as shown in Fig.~\ref{fig:UEplot}, is monotonic then the inverse is also trivially true: the mean jet \pT would also decrease with increasing EA. In other words, the $Q^{2}$ distribution softens from low- to high-EA events. On the other hand, the semi-inclusive jet spectra are normalized by the number of triggers, which have a lower $Q^{2}$ cut off and therefore would not be modified by an EA-to-$Q^{2}$ correlation at the same level as the jets. Consequently, an apparent suppression of the semi-inclusive jet spectrum in high-EA events compared with low-EA events could occur without any jet quenching. Furthermore, if the EA-to-$Q^{2}$ correlation is the sole cause, the suppression levels for trigger- and recoil-side jet spectra would be the same, as is the case in Fig.~\ref{fig:semi_inc_jets}.

\subsection{Dijet Azimuthal Difference and Momentum Imbalance Distributions} \label{sec:Aphi}

\begin{figure}[t!] \centering
    \includegraphics[trim=0.2cm 2.9cm 0.1cm 0.1cm, clip, width=0.45\textwidth]
    {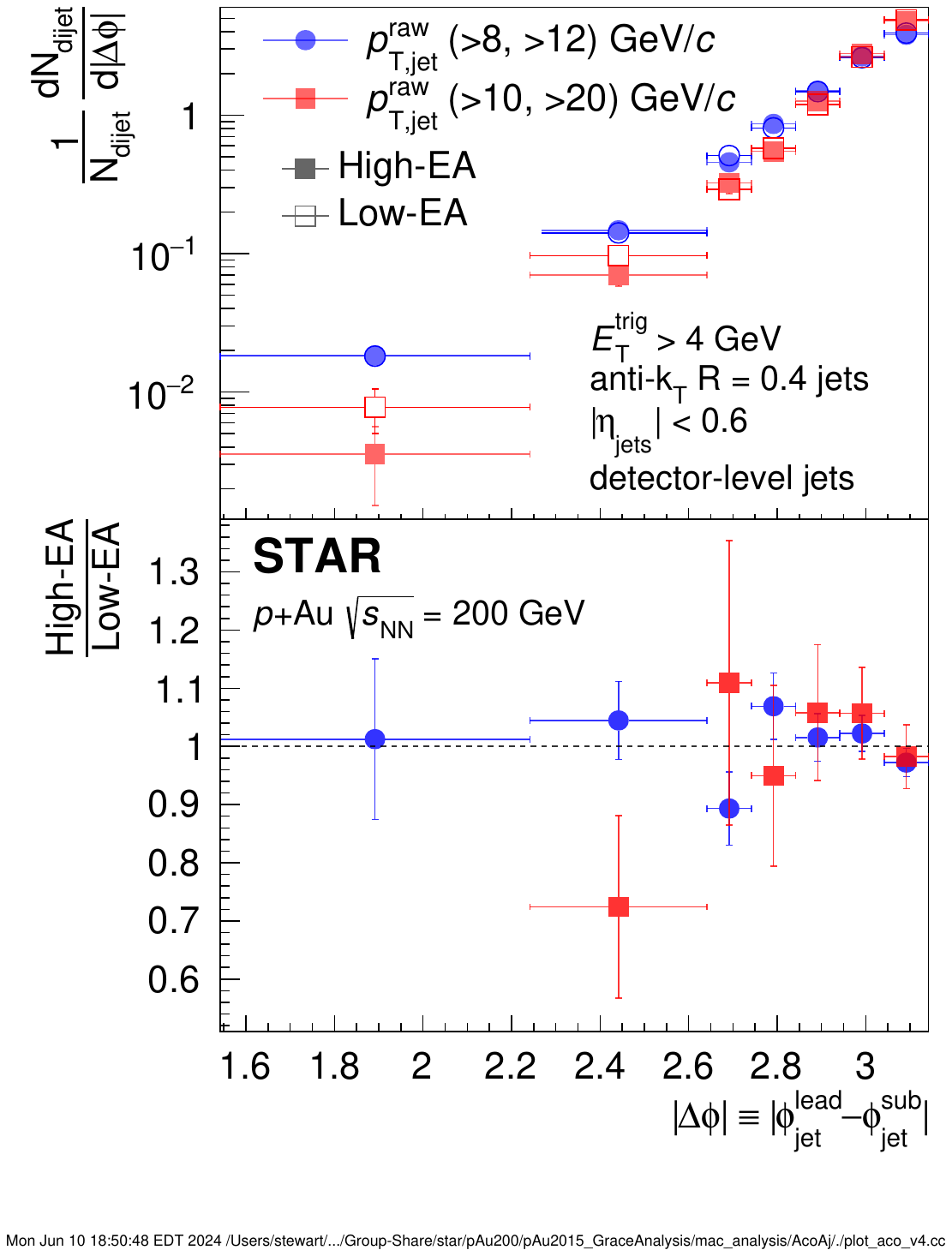}
    \caption{ %
        Top panel: distributions of the azimuthal separation between the two highest-\pT jets for high- and low-EA events. The distributions for the lower (higher) $p_\mathrm{T}^\mathrm{raw}$ requirements are shown in circles (squares).  Bottom panel: ratios between high-EA and low-EA events. All jets are detector-level jets.
    }\label{fig:Aphi}
\end{figure}

\begin{figure}[t!] \centering
    \includegraphics[trim=0.45 1.70cm 0.1cm 0.1cm, clip, width=0.45\textwidth]
    {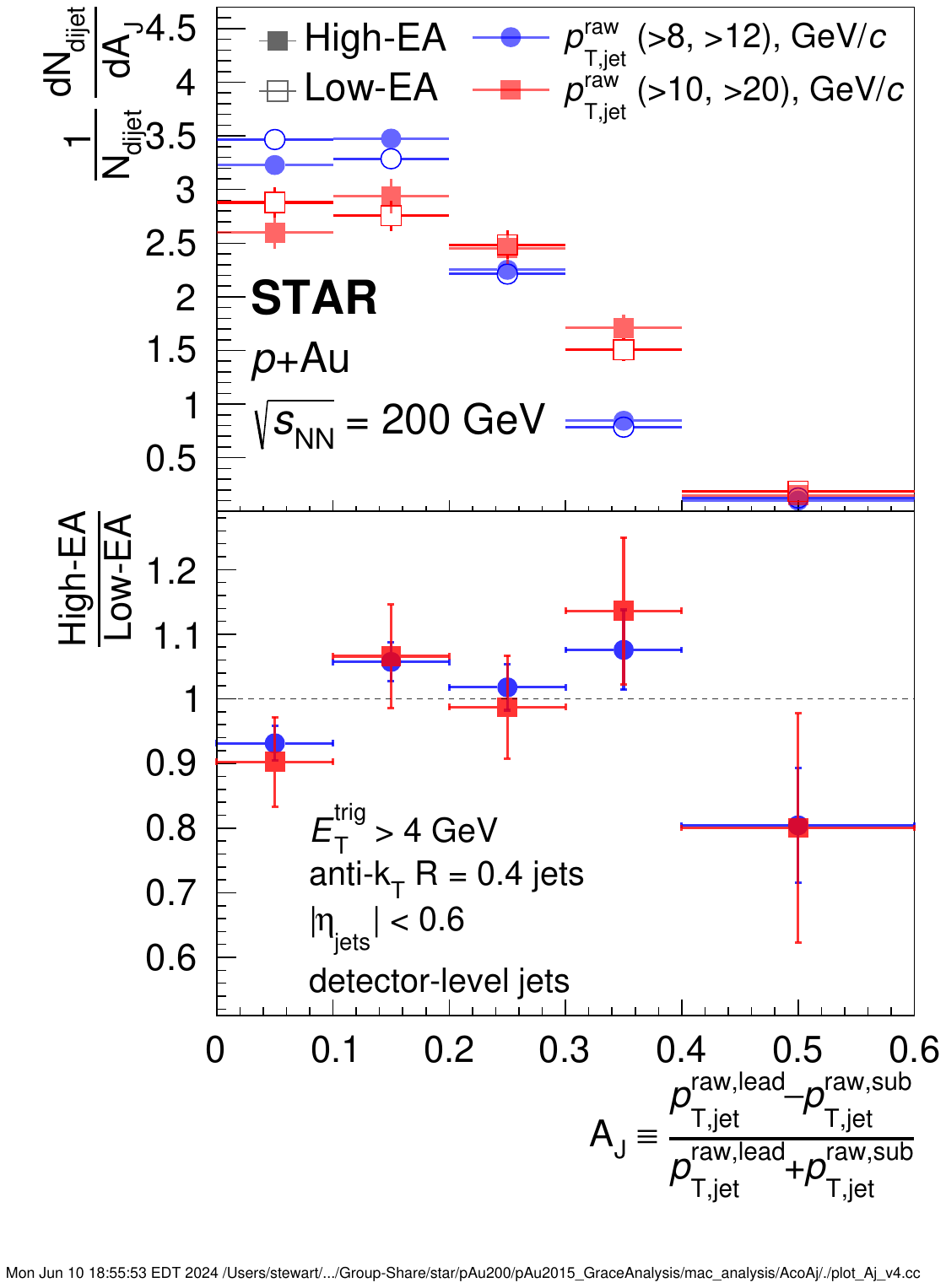}
    \caption{ %
        Top panel: distributions of dijet $p_\mathrm{T}^\mathrm{raw}$ imbalance, $A_\mathrm{J}$, for high- and low-EA events. The distributions for the lower (higher) $p_\mathrm{T}^\mathrm{raw}$ requirements are shown in circles (squares). Bottom panel: ratios between high-EA and low-EA events.  All jets are detector-level jets.
    }\label{fig:AJ}
\end{figure}

As previously mentioned, the EA-dependence of semi-inclusive jet spectra shown in Fig.~\ref{fig:semi_inc_jets} does not prove that the dependence results from jet quenching in high-EA events. To isolate a query for EA-dependent jet modification from the EA dependence of the $Q^{2}$ distribution, Figs.~\ref{fig:Aphi} and ~\ref{fig:AJ} present the detector-level per-dijet pair distributions of the dijet azimuthal difference (\Dphi) and momentum imbalance (\AJ) in high- and low-EA events. Because their normalization is per dijet, to first order the \Dphi and \AJ distributions avoid dependence on the underlying $p_\mathrm{T,jet}^\mathrm{raw}$ or \ETtrig distributions. %Therefore, even though \pTrawlead and \ETtrig distributions get softer at higher EA, the \AJ and \Dphi distributions would not necessarily be affected.

Dijets are selected according to the following criteria:

\begin{enumerate}
    \item $\ETtrig>\SI{4}{GeV}$
    \item Two different sets of thresholds for $p_\mathrm{T,jet}^\mathrm{raw,lead}$ and $p_\mathrm{T,jet}^\mathrm{raw,sub}$:
    \begin{itemize}
        \item  $p_\mathrm{T,jet}^\mathrm{raw,lead}$ ($p_\mathrm{T,jet}^\mathrm{raw,sub}$) $>$ 12  (8) $\mathrm{GeV}/c$
        \item $p_\mathrm{T,jet}^\mathrm{raw,lead}$ ($p_\mathrm{T,jet}^\mathrm{raw,sub}$) $>$ 20 (10) $\mathrm{GeV}/c$ 
    \end{itemize}
    \item Relative azimuth cut:
    \begin{itemize}
      \item For azimuthal difference: $|\Delta\phi|>\pi/2$
      \item For momentum imbalance: $|\Delta\phi|>(\pi-R_\mathrm{jet})$
   \end{itemize}
\end{enumerate}

While the measurements of \Dphi and \AJ are not corrected for detector effects and therefore not accessible to direct theoretical comparison, the detector effects cancel in the ratios between high- and low-EA events, which therefore can indicate the presence of jet quenching, if any.
As shown in Fig.~\ref{fig:Aphi}, there is no significant broadening of \Dphi for high-EA events relative to low-EA events, indicating that the leading and subleading jets retain their initial back-to-back configuration. Similarly, the dijet momentum imbalance distribution presented in Fig.~\ref{fig:AJ} shows no significant modification in high-EA events.

%%%%%%%%%%%%%%%%%%%%%%%%%%%%%%%%%%%%%%%
% Section: Conclusions                %
%%%%%%%%%%%%%%%%%%%%%%%%%%%%%%%%%%%%%%%
\section{Summary and Conclusion}\label{sec:Sum_and_Conc}

UE charged particle density is reported as a function of \ETtrig, \pTlead, $\eta$, and EA for \pAu collisions at \sNNcc. EA, measured at $\eta\in[-5,-3.4]$ in the BBC, and charged particle densities are positively correlated with a broad distribution, suggesting that EA measures event centrality with large fluctuations. 
At the lowest EA, the UE charged particle density is consistent with that in \pp collisions, and increases by about a factor of two in the highest EA events.

There is a statistically significant decrease in average EA with increasing \pTlead ($|\eta_\mathrm{jet}|<0.6$). Due to the large rapidity gap between the leading jet axis and the BBC, this anticorrelation must form at the initial stages of the collisions due to phase-space constraints between EA and $Q^2$ (or \xp) of the hard scattering.

The anticorrelation between EA and $Q^{2}$ suggests there would be an apparent suppression of high-\pT jet spectra at high-EA and independent of its physical cause when normalized by the number of triggers with a low threshold. This is demonstrated with the first measurements of semi-inclusive jet spectra for $p_\mathrm{T,jet}\in[15,40]$~GeV/$c$ and $\ETtrig>\SI{4}{GeV}$ in $p$+A collisions at RHIC. It is also notable that the suppression is similar for both the trigger- and recoil-side spectra; this would be expected if the suppression results from an \xp-to-EA bias, but might not be if resulting from path-length dependent jet quenching.

%% FIXME 

Finally, distributions of dijet azimuthal correlation and $p_\mathrm{T,jet}^\mathrm{raw}$ imbalance are reported, along with their ratios between high- and low-EA events. Because these distributions are normalized per dijet pair, they are to first order independent of any EA-to-$Q^2$ anticorrelation. Neither measurement shows significant EA dependence within uncertainties, indicating no sign of jet quenching in the high-EA \sNNcc $p$+Au collisions in this analysis. It should be noted that this result is for the top 30\% of EA events, and therefore does not directly compare with the top 5\% of EA events in $d$+Au collisions for which high-\pT $\pi^{0}$ production is reported to be suppressed relative to photon production \cite{PHENIX:2023dxl}.

Since the correlation between EA and jet production must stem from the earliest stages of the collisions \cite{Dumitru:2008wn}, \pAu collisions can serve as a potential probe for early-time dynamics or even the precollision configuration of the $p$ and/or gold nucleus.
These are promising opportunities for measurements with \pAu collisions that RHIC will provide in the near future, benefiting from the recently installed Event Plane Detector \cite{Adams:2019fpo} for better EA determination at large rapidity.

\begin{figure}[t!] \centering
    \includegraphics[width=0.48\textwidth, 
    trim=0.6cm 1.4cm 1.0cm 0.0cm, clip] {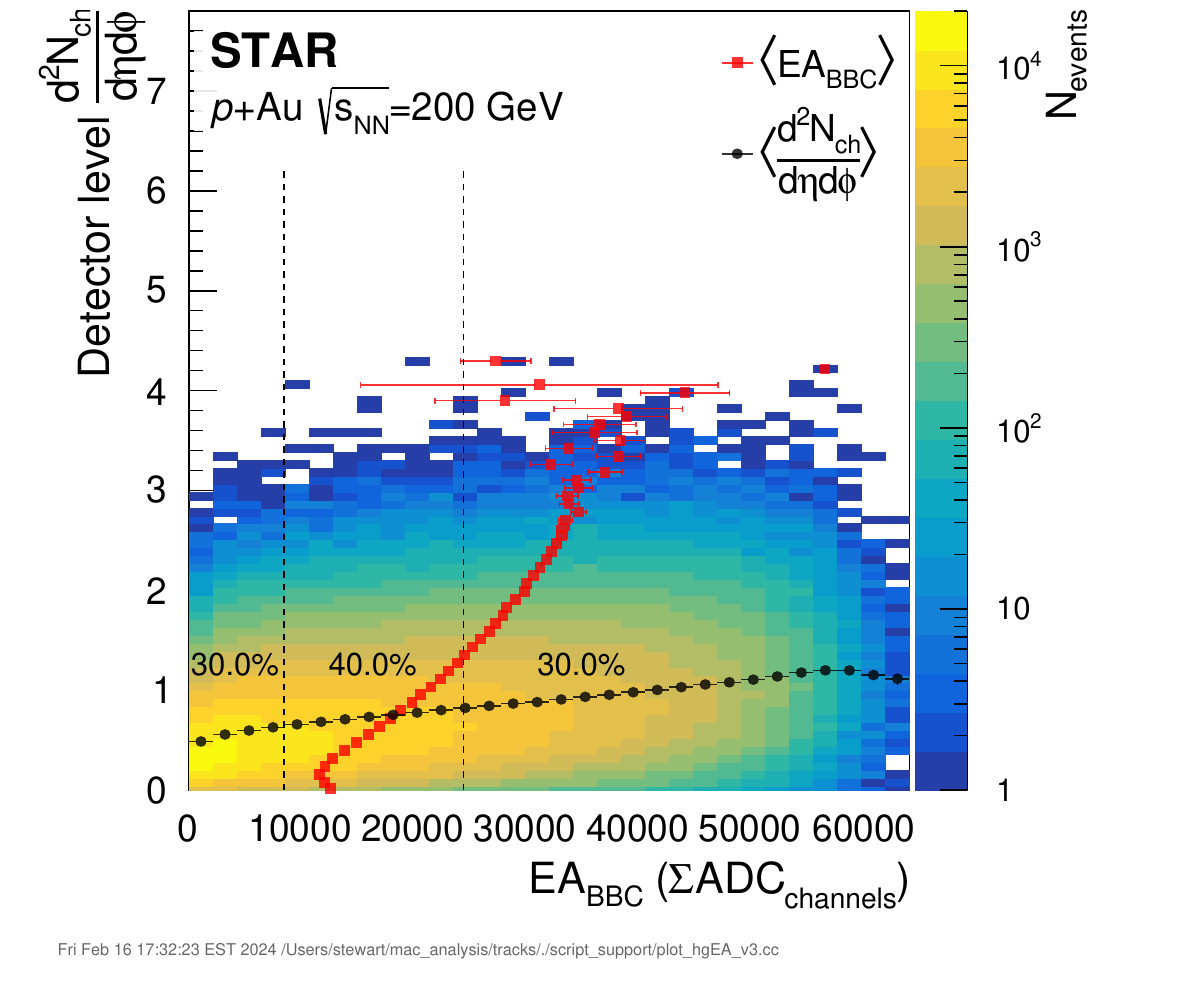} 
    \caption{Correlation of EA (summed ADC response from the 18 innermost BBC tiles) and uncorrected charged particle density in the TPC for MB events. The mean values of each distribution are plotted as a function of the other. Statistical errors on the means are shown, but are mostly smaller than the marker size.}
\label{fig:EA_vs_EA_MB} 
\end{figure}

%\newpage

\begin{figure}[ht!]
\centering
\includegraphics[width=0.48\textwidth, 
    trim=0.6cm 1.4cm 1.0cm 0.0cm, clip] {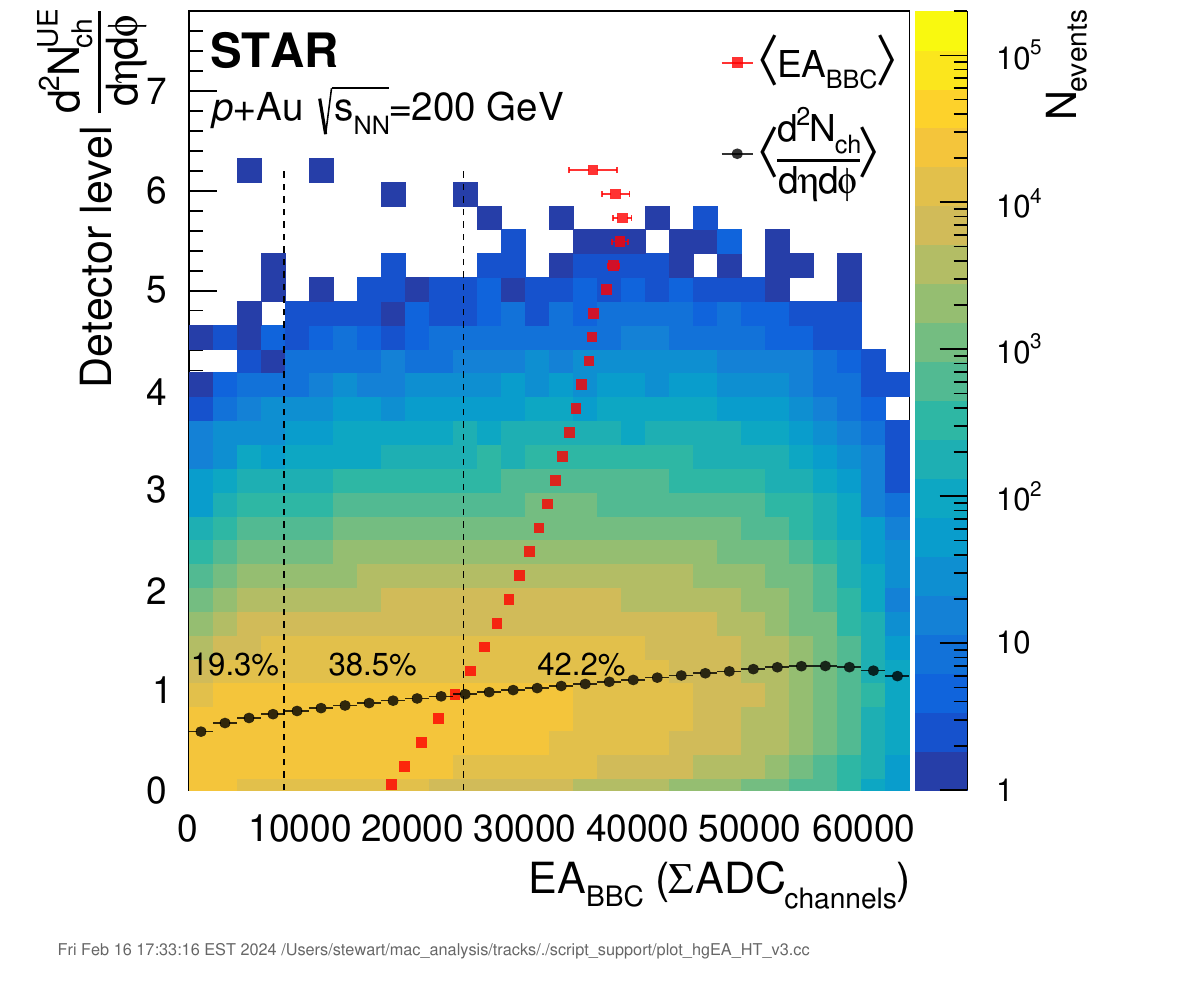} 
    \caption{ Correlation of EA (summed ADC response from the 18 innermost BBC tiles) and uncorrected charged particle density, for charged particles within $|\Delta\phi|\in[\pi/3,2\pi/3]$ relative to the trigger, in the TPC for HT events with an offline $\ET>\SI{4}{GeV}$ trigger requirement.  The mean values of each distribution are plotted as a function of the other. Statistical errors on the means are shown, but are mostly smaller than the marker size.}
\label{fig:EA_vs_EA_HT} 
\end{figure}

\section*{ACKNOWLEDGMENTS}

We thank the RHIC Operations Group and RCF at BNL, the NERSC Center at LBNL, and the Open Science Grid consortium for providing resources and support.  This work was supported in part by the Office of Nuclear Physics within the U.S. DOE Office of Science, the U.S. National Science Foundation, National Natural Science Foundation of China, Chinese Academy of Science, the Ministry of Science and Technology of China and the Chinese Ministry of Education, the Higher Education Sprout Project by Ministry of Education at NCKU, the National Research Foundation of Korea, Czech Science Foundation and Ministry of Education, Youth and Sports of the Czech Republic, Hungarian National Research, Development and Innovation Office, New National Excellency Programme of the Hungarian Ministry of Human Capacities, Department of Atomic Energy and Department of Science and Technology of the Government of India, the National Science Centre and WUT ID-UB of Poland, the Ministry of Science, Education and Sports of the Republic of Croatia, German Bundesministerium f\"ur Bildung, Wissenschaft, Forschung and Technologie (BMBF), Helmholtz Association, Ministry of Education, Culture, Sports, Science, and Technology (MEXT), Japan Society for the Promotion of Science (JSPS) and Agencia Nacional de Investigaci\'on y Desarrollo (ANID) of Chile.

%%%%%%%%%%%%%%%%%%%%%%%%%%%%%%%%%%%%%%%
% Appendix
%%%%%%%%%%%%%%%%%%%%%%%%%%%%%%%%%%%%%%%
\appendix

\section{Comparison of BBC Signal to Track Densities}\label{app:BBC_vs_trackDensity}

The relationship between EA and uncorrected charged charged particle density is given in \fig{EA_vs_EA_MB} for MB events and \fig{EA_vs_EA_HT} for HT triggered events with an offline cut of $E_\mathrm{T}^\mathrm{trig}>\SI{4}{GeV}$. The mean values of the charged particle densities at each EA selection, and conversely EA at each charged particle density selection, are also plotted. The statistical errors on the means are shown but are smaller than the marker size except in the tails of the distributions. 
The percentages of events within three ranges of EA are also plotted in Figs.~\ref{fig:EA_vs_EA_MB} and \ref{fig:EA_vs_EA_HT}; the lowest and highest 30\% in the MB sample define the ranges of low- and high-EA, respectively (see Sec.~\ref{sec:EA}).

\newpage
\bibliographystyle{apsrev4-2}
\bibliography{inspire}
\end{document}